\documentclass[a4paper,10pt,twoside]{cpc-hepnp}

\usepackage{multicol}
\usepackage{graphicx}
\usepackage{booktabs}
\usepackage{amssymb,bm,mathrsfs,bbm,amscd}
\usepackage[tbtags]{amsmath}
\usepackage{lastpage}

\begin{document}

\fancyhead[co]{\footnotesize L. Tolos et al. : Charmed hadrons in nuclear medium}

\footnotetext[0]{Received... }

\title{Charmed hadrons in nuclear medium}

\author{%
      L. Tolos$^{1;1)}$\email{tolos@kvi.nl}%
\quad  D. Gamermann$^{2}$ 
\quad  C. Garcia-Recio$^3$
\quad  R. Molina$^2$ 
\quad  J. Nieves$^2$
\quad  E. Oset$^2$ 
\quad  A. Ramos$^4$
}
\maketitle

\address{%
1~(Theory Group. KVI. University of Groningen, \\
Zernikelaan 25, 9747 AA Groningen, The Netherlands)\\
2~(Instituto de F{\'\i}sica Corpuscular (centro mixto CSIC-UV)\\
Institutos de Investigaci\'on de Paterna, Aptdo. 22085, 46071, Valencia, Spain)\\
3~(Departamento de F{\'\i}sica At\'omica, Molecular y Nuclear, \\
Universidad de Granada, E-18071 Granada, Spain)\\
4~(Departament d'Estructura i Constituents de la Mat\`eria,\\
Universitat de Barcelona,
Diagonal 647, 08028 Barcelona, Spain)\\
}

\begin{abstract}
We study the properties of charmed hadrons in dense matter within a  coupled-channel approach which accounts for Pauli blocking effects and meson self-energies in a self-consistent manner. We analyze the behaviour in this dense environment of dynamically-generated baryonic resonances as well as the open-charm meson spectral functions. We discuss the implications of the in-medium properties of open-charm mesons on the  $D_{s0}(2317)$ and the predicted $X(3700)$ scalar resonances.
\end{abstract}

\begin{keyword}
open-charm mesons, spectral function, dynamically-generated baryonic resonances, charmed and hidden charmed scalar resonances
\end{keyword}

\begin{pacs}
14.20.Lq, 11.10.St, 12.38.Lg, 14.40.Lb
\end{pacs}

\begin{multicols}{2}

\section{Introduction}
The future CBM (Compressed Baryonic Matter) experiment of the FAIR (Facility of Antiproton and Ion Research) project at GSI will investigate highly compressed dense matter in nuclear collisions  \cite{gsi}. An important part of the hadron physics project is devoted to the in-medium modification of hadrons in the charm sector and to provide a first insight into charm-nucleus interaction. Therefore, the modifications of the properties of open and hidden charmed mesons in a hot and dense environment are being the focuss of recent studies.

The in-medium modification of the properties of open-charm mesons ($D$ and $\bar D)$ may help to explain the $J/\Psi$ suppression in a hadronic environment as well as the possible formation of $D$-mesic nuclei. Moreover, changes in the properties of open-charm mesons will affect the renormalization of charmed and hidden charmed scalar meson resonances in nuclear matter, providing information about their nature, whether they are better described as $q\bar{q}$ states or dynamically generated resonances or a mixture of both schemes.

In the present work we present a study of the properties of open-charm mesons in dense matter within a self-consistent approach in coupled channels. We analyze the behaviour of dynamically generated charmed baryonic resonances as well as the open-charm meson spectral functions in this dense medium. We then analyze the effect of the self-energy of $D$ mesons on the properties of dynamically-generated charmed and hidden charmed scalar resonances, such as the $D_{s0}(2317)$ and the predicted $X(3700)$ resonances.

\section{Open charm in nuclear matter}
The self-energy and, hence, the spectral function for open-charm ($D$ and $\bar D$) mesons is obtained following a self-consistent coupled-channel procedure. The kernel of the Bethe-Salpeter equation or $T$-matrix ($T$) results from a transition potential coming from effective lagrangians. We will discuss two possible approaches to this bare interaction in the following. The self-energy is then obtained summing the transition amplitude $T$ for the different isospins over the nucleon Fermi distribution at a given temperature, $n(\vec{q},T)$, as 
\begin{eqnarray}
&&\Pi(q_0,{\vec q},T)= \int \frac{d^3p}{(2\pi)^3}\, n(\vec{p},T) \,  \nonumber \\
&& \times \, [\, {T}^{(I=0)} (P_0,\vec{P},T) +
3 \, {T}^{(I=1)} (P_0,\vec{P},T)\, ]\ , \label{eq:selfd}
\end{eqnarray}

%%%%%%%%%%%%%%%%%%%%%%%%%%%%%%%%%%%%%%%%%%%%%%%%%%%%%%%%%%%%%%%%%%%%%%%%%%%%5
\begin{center}
\includegraphics[width=0.45\textwidth, height=7cm]{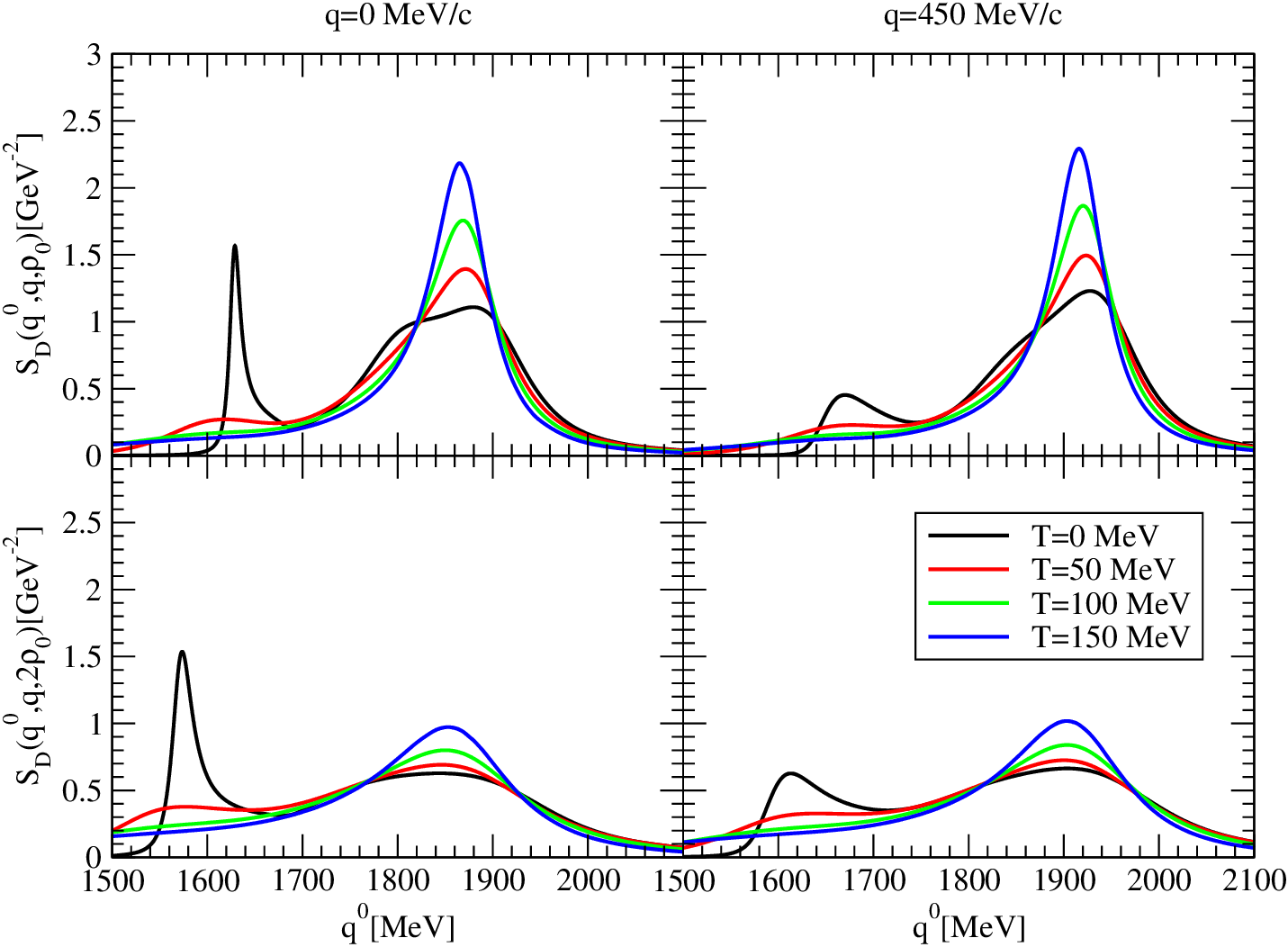}
\figcaption{The $D$ meson spectral function for different momenta, temperatures and densities for the SU(4) model \label{fig1}}
\end{center}
%%%%%%%%%%%%%%%%%%%%%%%%%%%%%%%%%%%%%%%%%%%%%%%%%%%%%%%%%%%%%%%%%%%%%%%%%%%%%

\noindent
where $P_0=q_0+E_N(\vec{p},T)$ and $\vec{P}=\vec{q}+\vec{p}$ are
the total energy and momentum of the meson-nucleon pair in the nuclear
matter rest frame, and ($q_0$,$\vec{q}\,$) and ($E_N$,$\vec{p}$\,) stand  for
the energy and momentum of the meson and nucleon, respectively, also in this
frame. The self-energy must be determined self-consistently since it is obtained from the
in-medium amplitude $T$ which contains the meson-baryon loop function, and this last quantity itself
is a function of the self-energy. The meson spectral function then reads
\begin{eqnarray}
S(q_0,{\vec q}, T)= -\frac{1}{\pi}\frac{{\rm Im}\, \Pi(q_0,\vec{q},T)}{\mid
q_0^2-\vec{q}\,^2-m^2- \Pi(q_0,\vec{q},T) \mid^2 } \ .
\label{eq:spec}
\end{eqnarray}

\subsection{SU(4) $t$-vector meson exchange models}
\label{su4}

The open-charm meson spectral functions are obtained from the multichannel Bethe-Salpeter equation taking, as bare interaction, a type of broken $SU(4)$ $s$-wave Weinberg-Tomozawa (WT)  interaction supplemented by an attractive isoscalar-scalar term and using a cutoff regularization scheme. This cutoff is fixed by generating dynamically the $I=0$ $\Lambda_c(2593)$ resonance. As a result, a new resonance in $I=1$ channel $\Sigma_c(2880)$ is generated \cite{LUT06,mizutani}. The in-medium solution at finite temperature 
incorporates  Pauli blocking effects, baryon mean-field bindings and $\pi$ and $D$ meson self-energies \cite{TOL07}.

In  Fig.~\ref{fig1} we display the $D$ meson spectral function for different momenta, densities and temperatures. At $T=0$ the spectral function shows two peaks. The $\tilde \Lambda_c N^{-1}$ excitation is seen at a lower energy whereas the second one at higher energy corresponds to the quasi(D)-particle  peak  mixed with  the $\tilde \Sigma_c N^{-1}$ state.  Those structures dilute with increasing temperature while the quasiparticle peak gets closer to its free value becoming narrower, as the self-energy
receives contributions from higher momentum $DN$ pairs where the interaction is weaker.
Finite density results in a broadening of the spectral function because of the increased  phase space. Similar effects were observed previously for the $\bar K$ in hot dense nuclear matter \cite{Tolos:2008di}.

\subsection{SU(8) model with \\ heavy-quark symmetry}

Heavy-quark symmetry (HQS) is a QCD spin-flavor symmetry
that appears when the quark masses, such as the charm mass, become
larger than the typical confinement scale. Then, the spin interactions vanish for infinitely massive
quarks. Thus, heavy hadrons come in doublets (if the spin of the light
degrees of freedom is not zero), which are degenerate in the infinite
quark-mass limit. And this is the case for the $D$ meson and its
vector partner, the $D^*$ meson. 

Therefore, we calculate the self-energy and, hence, the spectral function of the $D$ and $D^*$ mesons in nuclear matter simultaneously from a self-consistent calculation in coupled channels. To incorporate HQS to the meson-baryon interaction
we extend the WT meson-baryon lagrangian to the $SU(8)$ spin-flavor
symmetry group as we include pseudoscalars and vector mesons together with $J=1/2^+$ and $J=3/2^+$ baryons \cite{magas}, following the steps for $SU(6)$ of Ref.\cite{GarciaRecio:2005hy}. However, the $SU(8)$ spin-flavor is strongly broken in nature. On one hand, we  take into account mass
breaking effects by adopting the physical hadron masses in the tree
level interactions and in the evaluation of the
kinematical thresholds of different channels, as done in the previous $SU(4)$ models. On the other hand, we consider the difference between the weak non-charmed and charmed
pseudoscalar and vector meson decay constants. We also improve on the regularization scheme in nuclear matter going beyond the usual cutoff scheme \cite{tolos09}.

The $SU(8)$ model generates a wider spectrum of resonances with charm $C=1$
and strangeness $S=0$  compared to the  $SU(4)$ models, as seen in  Fig.~\ref{fig2}.
 While the parameters of both $SU(4)$ and $SU(8)$ models are fixed by the ($I=0$,$J=1/2$)
$\Lambda_c(2595)$ resonance, the fact that we incorporate vectors mesons in
the $SU(8)$ scheme generates naturally $J=3/2$ resonances, such as
$\Lambda_c(2660)$, $\Lambda_c(2941)$, $\Sigma_c(2554)$ and
$\Sigma_c(2902)$, some of which might be identified experimentally
\cite{Amsler}. 
%%%%%%%%%%%%%%%%%%%%%%%%%%%%%%%
\begin{center}
\includegraphics[width=0.45\textwidth, height=8.5cm]{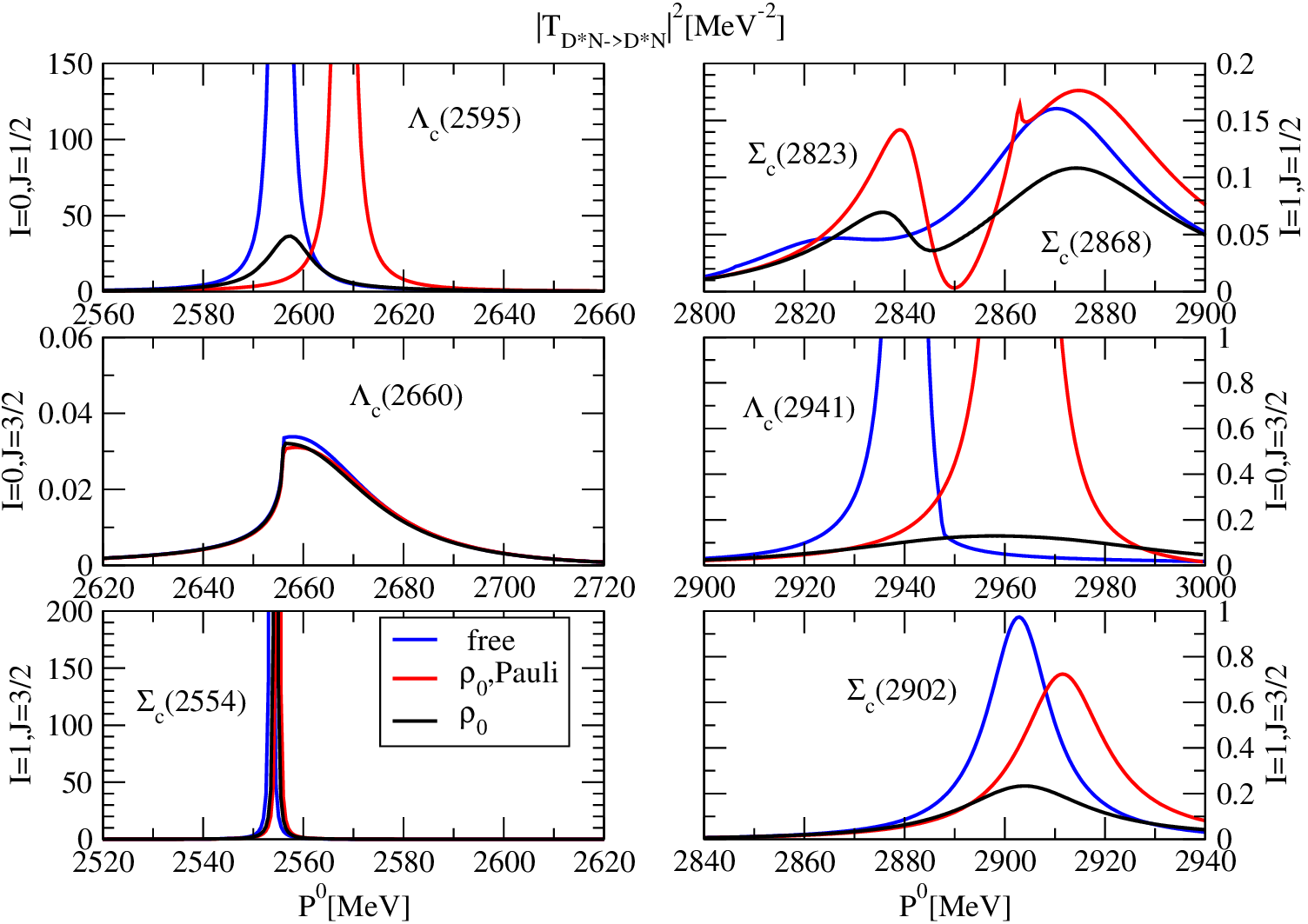}
\figcaption{Dynamically-generated charmed baryonic resonances in nuclear matter in the SU(8) scheme \label{fig2}}
\end{center}
%%%%%%%%%%%%%%%%%%%%%%%%%%%%%%%%%
\noindent
New resonances are also produced for $J=1/2$, as
$\Sigma_c(2823)$ and $\Sigma_c(2868)$, while others are not observed in $SU(4)$ models because of the different symmetry breaking pattern used in both models.

%%%%%%%%%%%%%%%%%%%%%%%%%%%%%%%%
\begin{center}
\includegraphics[width=0.45\textwidth, height=6cm]{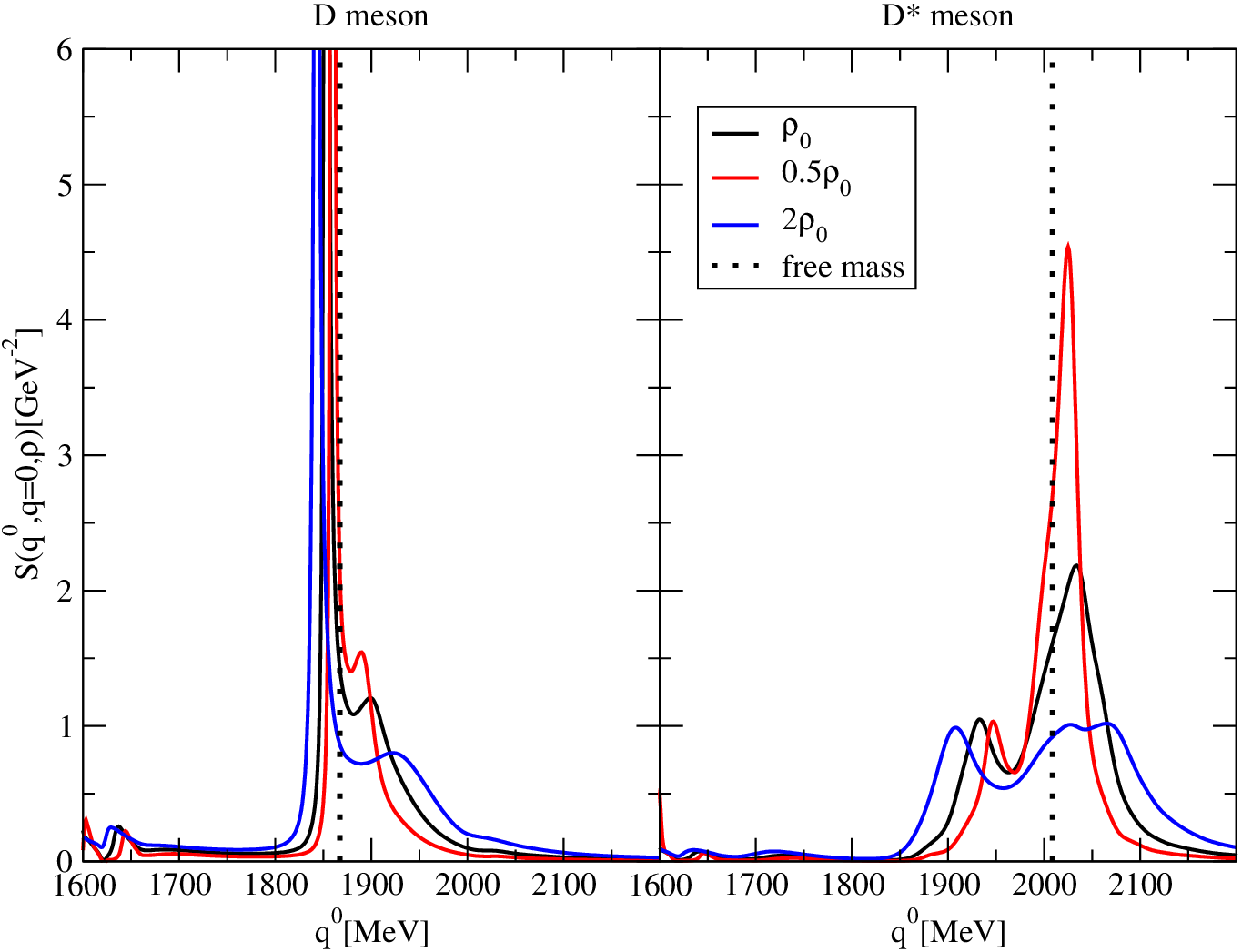}
\figcaption{$D$ and $D^*$ spectral functions in nuclear matter at $q=0$ MeV/c in the SU(8) scheme \label{fig3}}
\end{center}

%%%%%%%%%%%%%%%%%%%%%%%%%%%%%%%%%%%%%

The modifications of the mass and width of these resonances in
the nuclear medium will strongly depend on the coupling to channels
with $D$, $D^*$ and nucleon content, which are modified in the nuclear medium. Moreover, the resonances close to
the $DN$ or $D^*N$ thresholds change their properties more evidently
as compared to those far offshell. The improvement in the
regularization/renormalization procedure of the intermediate
propagators in the nuclear medium beyond the usual cutoff method has
also an important effect on the in-medium changes of the
dynamically-generated resonances, in particular, for those lying far
offshell from their dominant channel, as the case of the
$\Lambda_c(2595)$.

In  Fig.~\ref{fig3} we display the $D$ and $D^*$ spectral functions, which show then a rich spectrum of resonant-hole states. The $D$
meson quasiparticle peak mixes strongly with $\Sigma_c(2823)N^{-1}$
and $\Sigma_c(2868)N^{-1}$ states while the $\Lambda_c(2595)N^{-1}$ is
clearly visible in the low-energy tail. The $D^*$ spectral function
incorporates the $J=3/2$ resonances, and the quasiparticle peak fully mixes with $\Sigma_c(2902)N^{-1}$ and $\Lambda_c(2941)N^{-1}$. 
As
density increases, these $Y_cN^{-1}$ modes tend to smear out and the
spectral functions broaden with increasing phase space.

\section{Charmed and hidden charmed \\ scalar resonances in nuclear matter}

The analysis of the properties of scalar resonances in nuclear matter is a valuable tool in order to understand the nature of those states, whether they are $q \bar q$, molecules, mixtures of $q \bar q$ with meson-meson components, or dynamically generated resonances resulting from the interaction of two pseudoscalars. 

We study the charmed resonance $D_{s0}(2317)$ \cite{Kolomeitsev:2003ac,guo,Gamermann:2006nm} together with a hidden charmed scalar meson, $X(3700)$, predicted in Ref.~\cite{Gamermann:2006nm}, which might have been observed by the Belle collaboration \cite{Abe:2007sy} via the reanalysis of Ref.~\cite{Gamermann:2007mu}. Those resonances are generated dynamically solving the coupled-channel Bethe-Salpeter equation for two pseudoscalars \cite{Molina:2008nh}. The kernel is derived from a $SU(4)$ extension of the $SU(3)$ chiral Lagrangian used to generate scalar resonances in the light sector. The $SU(4)$ symmetry is, however, strongly 
 broken, mostly due to the explicit consideration of the masses of the vector 
 mesons exchanged between pseudoscalars \cite{Gamermann:2006nm}. 

The transition amplitude around each resonance for the different coupled channels gives us information about the coupling of this state to a particular channel. The $D_{s0}(2317)$ mainly couples to the $DK$ system, while the hidden charmed state $X(3700)$ couples most strongly to $D\bar{D}$. Then, any change in the $D$ meson properties in nuclear matter will have an important effect on these  resonances. Those modifications are given by the $D$ meson self-energy in the $SU(4)$ model without the phenomenological isoscalar-scalar term, but supplemented by the $p$-wave self-energy through the corresponding $Y_cN^{-1}$ excitations \cite{Molina:2008nh}.

%%%%%%%%%%%%%%%%%%%%%%%%%%%%%%%
\begin{center}
\includegraphics[width=0.4\textwidth,height=5cm]{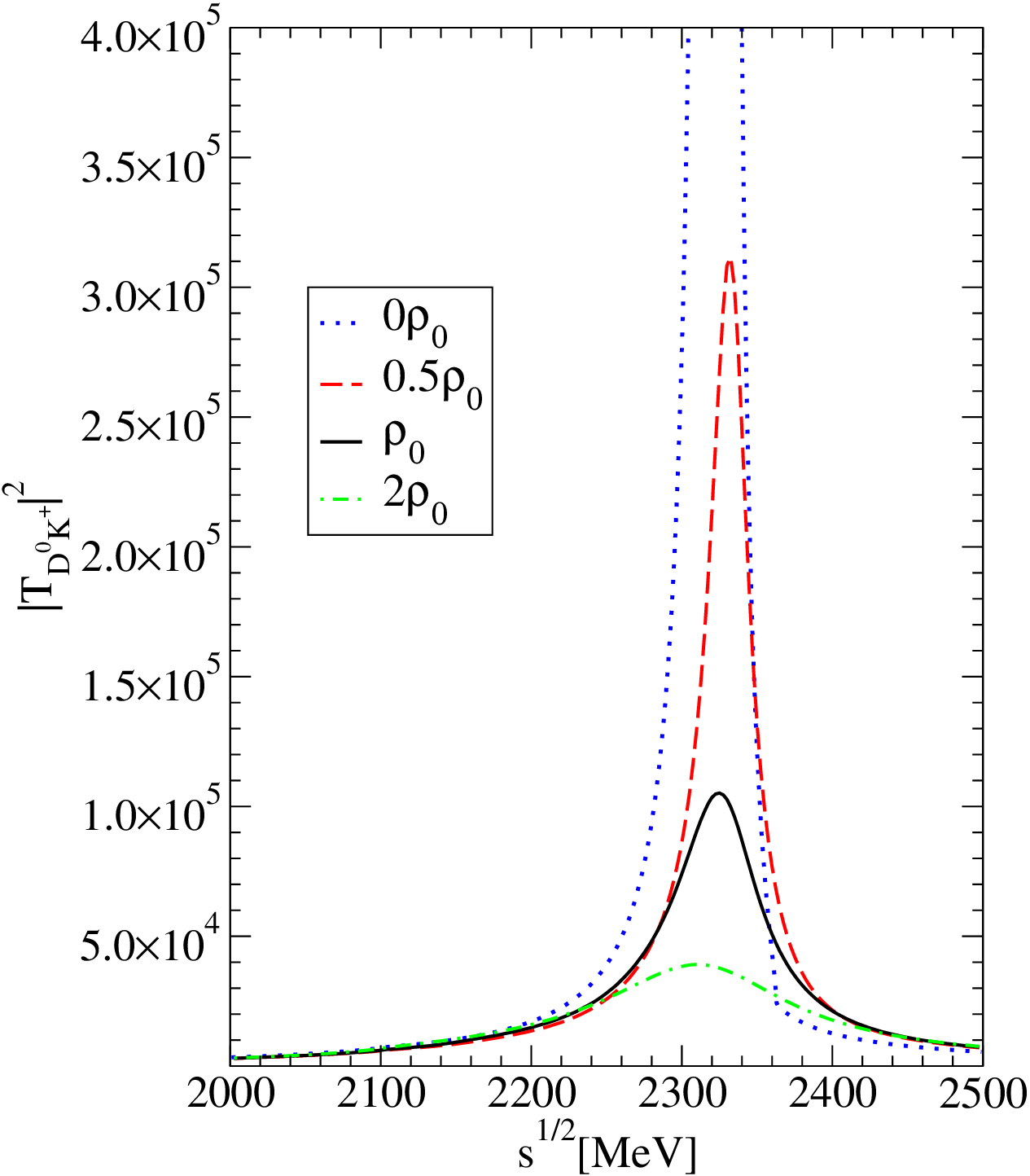}
\figcaption{$D_{s0}(2317)$ resonance in nuclear matter \label{fig4}}
\end{center}
%%%%%%%%%%%%%%%%%%%%%%%%%%%%%%%
%%%%%%%%%%%%%%%%%%%%%%%%%%%%%%
\begin{center}
\includegraphics[width=0.4\textwidth,height=5cm]{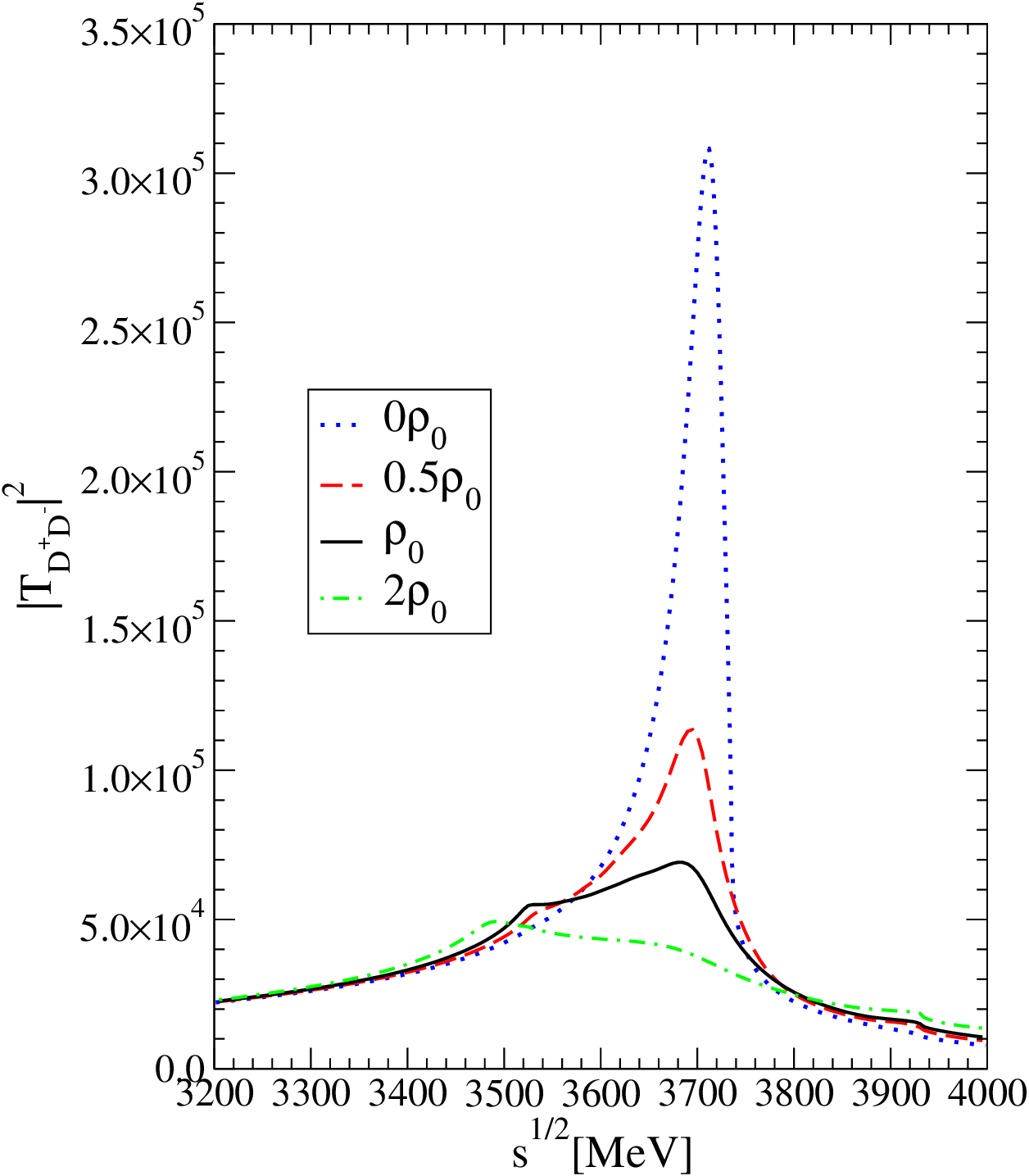}
\figcaption{$X(3700)$ resonance in nuclear matter \label{fig5}}
\end{center}

%%%%%%%%%%%%%%%%%%%%%%%%%%%%%%%%%%%%%5

 In Figs.~\ref{fig4} and \ref{fig5} the resonances $D_{s0}(2317)$ and $X(3700)$ are shown by displaying the squared transition amplitude for the corresponding dominant channel at different densities. The $D_{s0}(2317)$ and 
$X(3700)$ resonances, which have a zero and small width,
develop widths of the order of 100 and 200
MeV at normal nuclear matter density, respectively. The origin can be traced back to the opening of new many-body decay channels, as the $D$ meson gets absorbed in the nuclear medium via $DN$ and $DNN$ inelastic reactions. As for a mass shift, we do not extract any clear conclusion. We suggest to look at transparency ratios to investigate those in-medium widths. This magnitude, which gives the survival probability in production reactions in  nuclei, is very sensitive to the absorption rate of any resonance inside nuclei, i.e., to its in-medium width.

\section{Conclusions and Outlook}

We have studied the properties of open-charm mesons in dense matter within a self-consistent coupled-channel approach taking, as bare interaction, different effective lagrangians. The in-medium solution  accounts for Pauli blocking effects and meson self-energies. We have analyzed the behaviour in this dense environment of dynamically-generated charmed baryonic resonances together with the evolution with density and temperature of the open-charm meson spectral functions. We have finally discussed the implications of the properties of charmed mesons on the  $D_{s0}(2317)$ and the predicted $X(3700)$ in nuclear matter. We suggest to look at transparency ratios to investigate the changes in width of those resonances in nuclear matter.

\acknowledgments{L.T. acknowledges support from the RFF program of the University of Groningen. This work is partly supported by the EU contract No. MRTN-CT-2006-035482 (FLAVIAnet), by the contracts FIS2008-01661 and FIS2008-01143 from MICINN (Spain), by the Spanish Consolider-Ingenio 2010 Programme CPAN (CSD2007-00042), by the Generalitat de Catalunya contract 2009SGR-1289 and by Junta de Andaluc\'{\i}a under contract FQM225. We acknowledge the support of the European Community-Research Infrastructure Integrating Activity ``Study of Strongly Interacting Matter'' (HadronPhysics2, Grant Agreement n. 227431) under the 7th Framework Programme of EU.}

\end{multicols}

\vspace{-2mm}
\centerline{\rule{80mm}{0.1pt}}
\vspace{2mm}

\begin{multicols}{2}

\end{multicols}

\end{document}